\documentclass[12pt]{amsart}
\usepackage{amsfonts,amssymb}
\RequirePackage{ifpdf}
\usepackage{amsmath}
\usepackage{color}
\usepackage{pstcol,pst-node}
\usepackage{graphics}
\usepackage{epic}
\usepackage{curves}

\def\appendix#1{
\addtocounter{section}{1} \setcounter{equation}{0}
\renewcommand{\thesection}{\Alph{section}}
\section*{Appendix \thesection\protect\indent\quad
#1}
}
\renewcommand{\theequation}{\thesection.\arabic{equation}}

\catcode`\@=11
\def\marginnote#1{}

\newcount\hour
\newcount\minute
\newtoks\amorpm
\hour=\time\divide\hour by60 \minute=\time{\multiply\hour by60
\global\advance\minute by-\hour}
\edef\standardtime{{\ifnum\hour<12 \global\amorpm={am}%
        \else\global\amorpm={pm}\advance\hour by-12 \fi
        \ifnum\hour=0 \hour=12 \fi
        \number\hour:\ifnum\minute<10 0\fi\number\minute\the\amorpm}}
\edef\militarytime{\number\hour:\ifnum\minute<100\fi\number\minute}

\def\draftlabel#1{{\@bsphack\if@filesw {\let\thepage\relax
      \xdef\@gtempa{\write\@auxout{\string
          \newlabel{#1}{{\@currentlabel}{\thepage}}}}}\@gtempa \if@nobreak
    \ifvmode\nobreak\fi\fi\fi\@esphack} \gdef\@eqnlabel{#1}}
    \def\@eqnlabel{}
\def\@vacuum{}
\def\draftmarginnote#1{\marginpar{\raggedright\scriptsize\tt#1}}

\def\draft{
  \oddsidemargin -.5truein
  \def\@oddfoot{\footnotesize \sl preliminary draft \hfil
    \rm\thepage\hfil\sl\today\quad\militarytime}
  \let\@evenfoot\@oddfoot \overfullrule 3pt
    \let\label=\draftlabel
    \let\marginnote=\draftmarginnote
  \def\@eqnnum{(\theequation)\rlap{\kern\marginparsep\tt\@eqnlabel}%
    \global\let\@eqnlabel\@vacuum}

  }
\voffset= - 0.5in \hoffset= - 1.0in

\newcommand{\tr}{\,{\rm Tr}\,}

\def\be{\begin{equation}}
\def\ee{\end{equation}}
\def\bea{\begin{eqnarray}}
\def\eea{\end{eqnarray}}
\def\<{\langle}
\def\>{\rangle}

\def\tr{\mathop{\rm{tr}}}

\def\ocomma{{\phantom{\Bigm|}^{\phantom {X}}_{\raise-1.5pt\hbox{,}}\!\!\!\!\!\!\otimes}}

\newtheorem{theorem}{Theorem}[section]
\newtheorem{lm}[theorem]{Lemma}

\theoremstyle{definition}
\newtheorem{example}[theorem]{Example}

\allowdisplaybreaks

\begin{document}

\title[The matrix model for hypergeometric Hurwitz numbers]
{The matrix model for  hypergeometric Hurwitz numbers}
\author{Jan Ambj{\o}rn$^\dagger$}\thanks{$^\dagger$Niels Bohr Institute,
Copenhagen University, Denmark, and IMAPP, Radboud University, Nijmengen, The Netherlands.
Email: ambjorn@nbi.dk.}
\author{Leonid Chekhov$^{\ast}$}\thanks{$^{\ast}$Steklov Mathematical
Institute and  Laboratoire Poncelet, Moscow, Russia; Center for Quantum Geometry of Moduli Spaces,
{\AA}rhus University,
Denmark. Email: chekhov@mi.ras.ru.}

\begin{abstract}
We present the multi-matrix models that are the generating functions for branched
covers of the complex projective line ramified over $n$ fixed points $z_i$, $i=1,\dots,n$,
(generalized Grotendieck's dessins d'enfants) of fixed genus, degree, and the ramification
profiles at two points, $z_1$ and $z_n$. We take a sum over all possible ramifications at other $n-2$ points
with the fixed length of the profile at $z_2$ and with the fixed total length of profiles at the
remaining $n-3$ points. All these models belong to a class of hypergeometric
Hurwitz models thus being tau functions of the Kadomtsev--Petviashvili (KP) hierarchy.
In the case described above, we can present the
obtained model as a chain of matrices with a (nonstandard) nearest-neighbor interaction 
of the type $\tr M_iM_{i+1}^{-1}$. We describe the
technique for evaluating spectral curves of such models, which opens the
possibility of applying the topological recursion for developing $1/N^2$-expansions of
these model. These spectral curves turn out to be of an algebraic type.
\end{abstract}

\maketitle

\section{Introduction}\label{s:intro}
\setcounter{equation}{0}

In general, Hurwitz numbers pertain to combinatorial classes of ramified mappings $f:{\mathbb CP^1\to\Sigma_g}$
of the complex projective line onto a Riemann surface of genus $g$. Commonly, single or double Hurwitz numbers
correspond to the cases in which ramification profiles (defined by the corresponding Young tableauxes
$\lambda$ or $\lambda$ and $\mu$) are respectively determined 
at one ($\infty$) or two ($\infty$ and $1$) distinct points
whereas we assume the existence of $m$ other distinct ramification points with only simple ramifications.

Generating functions for Hurwitz numbers have been considered for long in mathematical physics. Notably, Okounkov and
Pandharipande \cite{OP} had shown that the exponential of the generating function for
double Hurwitz numbers is a tau-function of the Kadomtsev--Petviashvili (KP) hierarchy. The same result was
obtained by A.~Yu.~Orlov and Shcherbin \cite{OS}, \cite{Or} using the Schur function technique and, in a more
general setting, by Goulden and Jackson \cite{GJ} using Plucker relations.

Orlov and Shcherbin \cite{OS} also addressed the case of the generating function for the case of
Grothendieck {\em dessins d'enfants} where we have only three ramification points with multiple
ramifications and the ramification profile is fixed at one or two of these points. In this case, they
have also concluded that the exponentials of the corresponding generating functions have to be tau functions
of the KP hierarchy. Actually, the results of \cite{OS} describe a wider class of generating functions
for {\em hypergeometric Hurwitz
numbers} (this term was coined there) in which we have a fixed number $n$ of ramification points in ${\mathbb C}P^1$
assuming that we fix profiles at two of these points and take a sum over profiles at all other points with weights
being proportional to the {\em lengths} of the remaining $n-2$ profiles. Recently, Harnad and Orlov \cite{HO}
showed that all these generating functions are in turn tau functions of the KP hierarchy.

The interest to Hurwitz numbers corresponding to Belyi pairs was revived by Zograf
\cite{Zograf} (see also \cite{KZ}) who provided recursion relations for the generating function
of Grothendieck's {\it dessins d'enfants} enumerating the Belyi pairs $(C,f)$, where $C$ is a smooth algebraic curve
and $f$ a meromorphic function $f:C\to \mathbb CP^1$ ramified only over the points $0,1,\infty\in \mathbb CP^1$.
In \cite{AC}, we proposed the matrix-model description of Belyi pairs, clean Belyi morphisms, and two-profile Belyi pairs 
thus showing that all these cases fall into the category of KP tau functions. Then
the multi-matrix-model representation for the hypergeometric Hurwitz numbers
was constructed in \cite{AMMN} but with a complicated interaction between matrices in the chain.
In the present note, we propose a more standard description of
hypergeometric Hurwitz numbers in the case where
we fix profiles at two ramification points, fix the length of the profile at the third point, 
and fix the total length of profiles at other $n-3$ points.

We recall some mathematical results relating Belyi pairs to Galois groups.

\begin{theorem}\label{thm:Belyi}{\rm (Belyi, \cite{Belyi})}  A smooth complex algebraic curve $C$ is defined
over the field of algebraic numbers $\overline{\mathbb Q}$ if and only if we have a nonconstant 
meromorphic function $f$ on $C$ $(f:C\to {\mathbb C}P^1)$ ramified only over the points $0,1,\infty\in {\mathbb C}P^1$.
\end{theorem}

For a Belyi pair $(C,f)$ let $g$ be the genus of $C$ and $d$ the degree of $f$. If we take the preimage
$f^{-1}([0,1])\subset C$ of the real line segment $[0,1]\in {\mathbb C}P^1$ we obtain a connected bipartite
 fat graph with $d$ edges with vertices being preimages of $0$ and $1$ and such that the cyclic ordering of edges entering
 a vertex comes from the orientation of the curve $C$. This led Grothendieck to formulating the following lemma.

\begin{lm}\label{lm:Grot} {\rm (Grothendieck, \cite{Grot})}
There is a one-to-one correspondence between the isomorphism classes of Belyi pairs and connected bipartite fat graphs.
\end{lm}

A Grothendieck {\it dessin d'enfant} is therefore a connected bipartite fat graph representing a Belyi pair.
It is well known that we can naturally extend the dessin $f^{-1}([0,1])\subset C$ corresponding to a Belyi pair
$(C,f)$ to a bipartite triangulation of the curve $C$. For this, we cut the complex plane along the (real) line
containing $0,1,\infty$ coloring upper half plane white and lower half plane gray. This defines the partition of
$C$ into white and grey triangles such that white triangles has common edges only with grey triangles. We then consider a dual graph in which edges are of three types.

\begin{figure}[tb]
%\hspace*{2cm}
%\epsfysize=6cm
%\vskip .2in
{\psset{unit=0.7}
\begin{pspicture}(-3,-2.4)(3,2.4)
\psframe[linecolor=white, fillstyle=solid, fillcolor=yellow](-3.7,0)(3.7,-2.5)
\pcline[linestyle=solid, linewidth=1pt](-3.7,0)(3.7,0)
%\pcline[linestyle=solid, linewidth=2pt](-1,0)(1,0)
\rput(-1,0){\pscircle*{.1}}
\rput(1,0){\pscircle*{.1}}
\psarc[linecolor=white, linestyle=solid, linewidth=10pt](-1,0){2}{60}{300}
\psarc[linecolor=red, linestyle=dashed, linewidth=1.5pt](-1,0){1.8}{60}{300}
\psarc[linecolor=blue, linestyle=solid, linewidth=1pt](-1,0){2.2}{60}{300}
\psarc[linecolor=green, linestyle=dotted, linewidth=2pt](-1,0){1.8}{-60}{60}
\psarc[linecolor=green, linestyle=dotted, linewidth=2pt](-1,0){2.2}{-60}{60}
\psarc[linecolor=white, linestyle=solid, linewidth=10pt](1,0){2}{-120}{120}
\psarc[linecolor=green, linestyle=dotted, linewidth=2pt](1,0){1.8}{-120}{120}
\psarc[linecolor=blue, linestyle=solid, linewidth=1pt](1,0){2.2}{-120}{120}
\psarc[linecolor=green, linestyle=dotted, linewidth=2pt](1,0){1.8}{120}{240}
\psarc[linecolor=red, linestyle=dashed, linewidth=1.5pt](1,0){2.2}{120}{240}
\pcline[linecolor=white, linestyle=solid, linewidth=10pt](0,1.73)(0,-1.73)
\pcline[linecolor=green, linestyle=dotted, linewidth=2pt](0.2,1.73)(0.2,-1.73)
\pcline[linecolor=green, linestyle=dotted, linewidth=2pt](-0.2,1.73)(-0.2,-1.73)
\pscircle[linecolor=white, linestyle=solid, linewidth=6pt](-1,0){2.15}
\pscircle[linecolor=white, linestyle=solid, linewidth=6pt](1,0){2.15}
\pcline[linecolor=white, linestyle=solid, linewidth=6pt](0,1.73)(0,-1.73)
\rput(0,1.73){\pscircle[linecolor=black, fillstyle=solid, fillcolor=white]{.25}}
\rput(0,-1.73){\pscircle*{.25}}
\rput(-3.7,0){\pscircle[linecolor=black, fillstyle=solid, fillcolor=white]{.1}}
\rput(3.7,0){\pscircle[linecolor=black, fillstyle=solid, fillcolor=white]{.1}}
\psframe[linecolor=white, fillstyle=solid, fillcolor=white](-3.9,0.2)(-3.7,-.2)
\psframe[linecolor=white, fillstyle=solid, fillcolor=white](3.9,0.2)(3.7,-.2)
\rput(-1.7,0){\pscircle[linecolor=black, fillstyle=solid, fillcolor=white]{.1}}
\rput(-.5,0){\pscircle[linecolor=black, fillstyle=solid, fillcolor=white]{.1}}
\rput(.5,0){\pscircle[linecolor=black, fillstyle=solid, fillcolor=white]{.1}}
\rput(1.7,0){\pscircle[linecolor=black, fillstyle=solid, fillcolor=white]{.1}}
\rput(-4.3,0){\makebox(0,0)[cc]{\hbox{{$\infty_-$}}}}
\rput(4.3,0){\makebox(0,0)[cc]{\hbox{{$\infty_+$}}}}
%\rput(-1,.4){\makebox(0,0)[cc]{\hbox{{$0$}}}}
%\rput(1,.4){\makebox(0,0)[cc]{\hbox{{$1$}}}}
\rput(1,1.5){\makebox(0,0)[cc]{\hbox{{\small$\Lambda$}}}}
\rput(1,-1.4){\makebox(0,0)[cc]{\hbox{{\small$\overline\Lambda$}}}}
\end{pspicture}
}
\caption{\small The generalized Belyi graph $\Gamma_1$ corresponding to
possible ramifications at $n=5$ points (commonly taken to be $\infty$, $-(1+\sqrt{5})/2$, $0$, $1$, and
$(3+\sqrt{5})/2$; we denote them by small white circles).
This graph describes the generalized Belyi pair $(\mathbb CP^1,\hbox{id})$; $\infty_{\pm}$
indicate directions of approaching the infinite point in $\mathbb CP^1$.
The symbols $\Lambda$ and $\overline \Lambda$ indicate the insertions of the external field in the matrix-model formalism
of Sec.~\ref{s:model}. For example, this graph contributes the term $N^2\gamma_1\gamma_2\gamma_3^2 t_1 \tr(\Lambda\overline\Lambda)$.}
\label{fi:Belyi}
\end{figure}
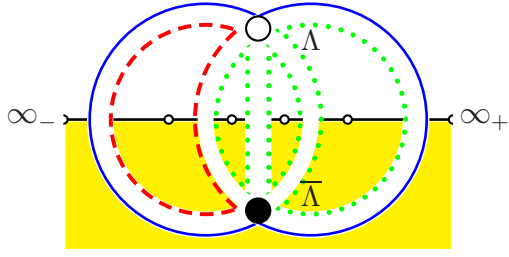

In this paper, we consider \emph{generalized Belyi pairs}, which are mappings $(f:C\to {\mathbb C}P^1)$ with possible
ramifications over $n$ fixed points $z_i\in {\mathbb C}P^1$, $i=1,\dots,n$. We then have the splitting of the
curve $C$ into bipartite $n$-gons with $n$ colored edges (the corresponding fat graphs are then coverings of the
basic graph depicted in Fig.~\ref{fi:Belyi} for $n=5$):
the type of an edge depend on which of $n$ segments of ${\mathbb R}P^1$---$f^{-1}([\infty_-,z_2])\subset C$,
$f^{-1}([z_2,z_3])\subset C$, $\dots$, $f^{-1}([z_{n-1},z_n])\subset C$, 
$f^{-1}([z_n,\infty_+])\subset C$---it intersects (we identify $z_1$ with the infinity point and let
$\infty_{\pm}$ indicate the directions of approaching this point
along the real axis in $\mathbb CP^1$).
Each face of the dual partition then contains a preimage of exactly
one of the points $z_1,\dots,z_n$, so these faces are of $n$ sorts 
(bordered by solid, dotted, or dashed lines in the figure). We call such a graph a {\it generalized Belyi fat graph}.

The type of ramification at infinity is determined by the set of solid-line-bounded faces of a generalized
Belyi fat graph: the order of branching is $r$ for a $2r$-gon, so we introduce the generating function that
distinguishes between different types of
branching at infinity, or $z_1$.
Moreover, we also distinguish between different types of ramifications at the
$n$th point (the point $(3+\sqrt{5})/2$ in Fig.~\ref{fi:Belyi}). This situation is often called a \emph{two-profile} generating function for
Hurwitz numbers because we fix two ramification patterns at two distinct branching points; each such pattern can be represented by a Young tableaux.
We let $k_i$ denote the numbers of respective cycles (pre-images of the points $z_i$ on the Riemann surface $C$)
and let $k_1^{(r)}$ and $k_n^{(r)}$ denote the
numbers of cycles of length $2r$ centered at pre-images of the respective points $z_1$ and $z_n$
in a generalized Belyi fat graph.

As was shown in \cite{AMMN} and \cite{HO}, the exponential of the generating function
\be
{\mathcal F}\bigl[\{t_m\},\{{\mathfrak t}_r\},\gamma_2,\dots,\gamma_{n-1};N\bigr]=\sum_{\Gamma}\frac{1}
{|\hbox{Aut\,}\Gamma|}N^{2-2g}\prod_{r=1}^{\infty}t_{r}^{k_1^{(r)}}
\prod_{s=1}^{\infty}{\mathfrak t}_{s}^{k_n^{(s)}}\prod_{j=2}^{n-1}\gamma_j^{k_j}
\label{gen-fun1}
\ee
is a tau function of the KP hierarchy in times $t$ or ${\mathfrak t}$. Although a matrix-model description of this
generating function was proposed in the above papers, the possibility of solving it in \emph{topological recursion} terms (see \cite{Ey}, \cite{ChEy}, \cite{CEO})
remained obscure. We are going to construct a matrix model describing a subclass of generating functions (\ref{gen-fun1}) with $\gamma_3=\gamma_4=\cdots=\gamma_{n-1}$ leaving $\gamma_2$ arbitrary.

Our goal in the present paper is therefore to construct and solve a matrix model whose free energy is the
generating function
\be
{\mathcal F}\bigl[\{t_m\},\{{\mathfrak t}_r\},\gamma_2,\gamma_{3};N\bigr]=\sum_{\Gamma}\frac{1}
{|\hbox{Aut\,}\Gamma|}N^{2-2g}\prod_{r=1}^{\infty}t_{r}^{k_1^{(r)}}
\prod_{s=1}^{\infty}{\mathfrak t}_{s}^{k_n^{(s)}}\gamma_2^{k_2}\gamma_3^{k_3+\cdots+k_{n-1}},
\label{gen-fun}
\ee
where $N$, $\gamma_2$, $\gamma_3$,
$t_r$, and ${\mathfrak t}_r$
are formal independent parameters and the sum ranges all (connected) generalized Belyi fat graphs. Below we
are dealing with a matrix model with an external matrix field
$\Lambda=\hbox{diag\,}(\lambda_1,\dots, \lambda_{\gamma_3N})$, the corresponding times are
\be
{\mathfrak t}_r=\tr\bigl[(\Lambda\overline\Lambda)^r\bigr].
\label{tt}
\ee

Sometimes factors $\gamma_1^{k_1}$ and $\gamma_n^{k_n}$ are added but they
can always be absorbed into the times $t_r$ and ${\mathfrak t}_r$
by scaling $t_r\to \gamma_1 t_r$ and ${\mathfrak t}_r\to \gamma_n {\mathfrak t}_r$
for all $r$.

The structure of the paper is as follows.
In Sec.~\ref{s:model}, we show that generating function (\ref{gen-fun}) is the free energy of a special multi-matrix
model represented as a chain of matrices with somewhat nonstandard interaction terms $\tr M_iM_{i+1}^{-1}$. We express this model
as an integral over eigenvalues of these matrices in a form similar to that of
the standard generalized Kontsevich model (GKM) \cite{MMM}. We adapt
the technique of Eynard and Prats Ferrer \cite{EPF} to evaluating spectral curves for chains of matrices 
with these nonstandard interaction terms in Sec.~\ref{s:CEO}. Although we derive the spectral curve only in the first
nontrivial case $n=4$ (the case of one intermediate field), our technique can be straightforwardly generalized to all higher $n$,
which will be done in a separate publication. 
We conclude with the discussion of
our results.

Throughout the entire text we disregard all multipliers not depending on external fields and times $t_r$; all equalities in the paper
must be therefore understood modulo such factors.

\section{The model}\label{s:model}
\setcounter{equation}{0}

In order to take into account the profile at the infinity point, we first contract all solid cycles (centered at
pre-images of $\infty$) assigning the time $t_r$ to every contracted cycle of length $2r$.

The new interaction vertices arise from the thus contracted
solid cycles. For example, for a cycle of
length four, we obtain the correspondence
$$
{\psset{unit=1}
\begin{pspicture}(-2,-3)(4,3)
\newcommand{\PATTERNTHREE}{%
{\psset{unit=1}
\pcline[linecolor=green, linestyle=dotted, linewidth=2pt](0,0.2)(1.2,0.2)
\pcline[linecolor=green, linestyle=dotted, linewidth=2pt](0,-.2)(1.2,-.2)
\rput{315}(0,0){
\pcline[linecolor=green, linestyle=dotted, linewidth=2pt](0,0.2)(1.2,0.2)
\pcline[linecolor=green, linestyle=dotted, linewidth=2pt](0,-.2)(1.2,-.2)
}
\rput{45}(0,0){
\pcline[linecolor=red, linestyle=dashed, linewidth=1.5pt](0,.2)(1.2,.2)
\pcline[linecolor=green, linestyle=dotted, linewidth=2pt](0,-.2)(1.2,-.2)
\pcline[linecolor=white, linestyle=solid, linewidth=6pt](0,0)(1.2,0)
}
\rput{315}(0,0){
\pcline[linecolor=white, linestyle=solid, linewidth=6pt](0,0)(1.2,0)
}
\pcline[linecolor=white, linestyle=solid, linewidth=6pt](0,0)(1.2,0)
}
}
\newcommand{\PATTERNTWO}{%
{\psset{unit=1}
\pcline[linecolor=green, linestyle=dotted, linewidth=2pt](0,0.2)(1.2,0.2)
\pcline[linecolor=green, linestyle=dotted, linewidth=2pt](0,-.2)(1.2,-.2)
\rput{315}(0,0){
\pcline[linecolor=green, linestyle=dotted, linewidth=2pt](0,0.2)(1.2,0.2)
\pcline[linecolor=red, linestyle=dashed, linewidth=1.5pt](0,-.2)(1.2,-.2)
}
\rput{45}(0,0){
\pcline[linecolor=green, linestyle=dotted, linewidth=2pt](0,0.2)(1.2,0.2)
\pcline[linecolor=green, linestyle=dotted, linewidth=2pt](0,-.2)(1.2,-.2)
\pcline[linecolor=white, linestyle=solid, linewidth=6pt](0,0)(1.2,0)
}
\rput{315}(0,0){
\pcline[linecolor=white, linestyle=solid, linewidth=6pt](0,0)(1.2,0)
}
\pcline[linecolor=white, linestyle=solid, linewidth=6pt](0,0)(1.2,0)
}
}
\psarc[linecolor=green, linestyle=dotted, linewidth=2pt](0,0){1.2}{55}{125}
\psarc[linecolor=green, linestyle=dotted, linewidth=2pt](0,0){1.2}{235}{305}
\psarc[linecolor=red, linestyle=dashed, linewidth=1.5pt](0,0){1.2}{145}{215}
\psarc[linecolor=red, linestyle=dashed, linewidth=1.5pt](0,0){1.2}{-35}{35}
\rput{45}(0,0){
\rput(1,0){\PATTERNTWO
}
}
\rput{225}(0,0){
\rput(1,0){
\PATTERNTWO
}
}
\rput{135}(0,0){
\rput(1,0){\PATTERNTHREE
}
}
\rput{315}(0,0){
\rput(1,0){\PATTERNTHREE
}
}
\pscircle[linecolor=white, fillstyle=solid, fillcolor=white](0,0){1.1}
\pscircle[linecolor=blue, linestyle=solid, linewidth=1pt](0,0){.8}
\rput(-0.15,1.3){\makebox(0,0)[rb]{\hbox{\small{$\Lambda$}}}}
\rput(0.15,1.3){\makebox(0,0)[lb]{\hbox{\small{$\overline\Lambda$}}}}
\rput(0.15,-1.3){\makebox(0,0)[lt]{\hbox{\small{$\Lambda$}}}}
\rput(-0.15,-1.3){\makebox(0,0)[rt]{\hbox{\small{$\overline\Lambda$}}}}
\rput(0,0){\makebox(0,0)[cc]{\hbox{{$\gamma_1$}}}}
\rput{135}(0,0){
\rput(1,0){\pscircle*{.2}}
}
\rput{315}(0,0){
\rput(1,0){\pscircle*{.2}}
}
\rput{45}(0,0){
\rput(1,0){\pscircle[linecolor=black, fillstyle=solid, fillcolor=white]{.2}}
}
\rput{225}(0,0){
\rput(1,0){\pscircle[linecolor=black, fillstyle=solid, fillcolor=white]{.2}}
}
\rput(2,.5){\makebox(0,0)[lb]{\hbox{\tiny{$\overline B_2$}}}}
\rput(2,-.5){\makebox(0,0)[lt]{\hbox{\tiny{$B_2$}}}}
\rput(-2,-.5){\makebox(0,0)[rt]{\hbox{\tiny{$\overline B_2$}}}}
\rput(-2,.5){\makebox(0,0)[rb]{\hbox{\tiny{$B_2$}}}}
\rput(.5,2){\makebox(0,0)[lb]{\hbox{\tiny{$\overline B_4$}}}}
\rput(.5,-2){\makebox(0,0)[lt]{\hbox{\tiny{$B_4$}}}}
\rput(-.5,-2){\makebox(0,0)[rt]{\hbox{\tiny{$\overline B_4$}}}}
\rput(-.5,2){\makebox(0,0)[rb]{\hbox{\tiny{$B_4$}}}}
\rput(1.5,1.5){\makebox(0,0)[lb]{\hbox{\tiny{$\overline B_3$}}}}
\rput(1.5,-1.5){\makebox(0,0)[lt]{\hbox{\tiny{$B_3$}}}}
\rput(-1.5,-1.5){\makebox(0,0)[rt]{\hbox{\tiny{$\overline B_3$}}}}
\rput(-1.5,1.5){\makebox(0,0)[rb]{\hbox{\tiny{$B_3$}}}}
\rput(2,0){\makebox(0,0)[lc]{\hbox{{$\sim \frac 12 N t_2\tr
\Bigl[\bigl(
B_2 B_3 B_4\Lambda \overline \Lambda\,{\overline B}_4 {\overline B}_3 {\overline B}_2\bigr)^2\Bigr]$}}}}
\end{pspicture}
}
$$
where the factor $1/2$ takes into account the symmetry of the four-cycle.

The matrix-valued fields $B_i$, $i=2,\dots,n-1$, are general complex-valued matrices such that
$B_2$ is a rectangular matrix of the size $\gamma_2N\times \gamma_3 N$ and we always assume that
$$
\gamma_2>\gamma_3,
$$
and all other matrices $B_3, \dots, B_{n-1}$ are square matrices of the size $\gamma_3N\times \gamma_3 N$.

The matrix-model integral whose free energy is the generating function (\ref{gen-fun}) reads
\be
\int DB_2\cdots DB_{n-1}e^{N\sum_{r=1}^\infty \frac{t_r}{r}\tr
\Bigl[\bigl(
B_2 \cdots B_{n-1}\Lambda \overline \Lambda\,{\overline B}_{n-1}\cdots {\overline B}_2\bigr)^r\Bigr]
-\sum_{j=2}^{n-1}N\tr (B_j\overline B_j)}
\label{model1}
\ee

We next perform the variable changing
\be
\begin{array}{l}
{\mathfrak B}_2=B_2B_3\cdots B_{n-1}\cr
{\mathfrak B}_3=B_3\cdots B_{n-1}\cr
\vdots\cr
{\mathfrak B}_{n-1}=B_{n-1}
\end{array}
\label{model2}
\ee
and assume that all matrices ${\mathfrak B}_3,\dots,{\mathfrak B}_{n-1}$ 
\emph{are invertible} (the matrix ${\mathfrak B}_2$
remains rectangular). With accounting for the Jacobian of transformation (\ref{model2}), the integral (\ref{model1})
becomes
\bea
&{}&\int D{\mathfrak B}_2\cdots D{\mathfrak B}_{n-1}\exp\Bigl\{
-\gamma_2N\tr\log({\mathfrak B}_3{\overline {\mathfrak B}}_3)-\sum_{j=4}^{n-1} \gamma_3N\tr\log({\mathfrak B}_j{\overline {\mathfrak B}}_j)\Bigr.\nonumber\\
&{}&+\sum_{r=1}^\infty N\frac{t_r}{r}\tr\Bigl[({\mathfrak B}_2|\Lambda|^2
{\overline {\mathfrak B}}_2)^r\Bigr]
-N\tr\bigl[{\mathfrak B}_2{\mathfrak B}^{-1}_3{\overline {\mathfrak B}}_3^{-1}
{\overline {\mathfrak B}}_2\bigr]\nonumber\\
&{}&\Bigl.
-N\tr\bigl[{\mathfrak B}_3{\mathfrak B}^{-1}_4{\overline {\mathfrak B}}_4^{-1}
{\overline {\mathfrak B}}_3\bigr]-\cdots
-N\tr\bigl[{\mathfrak B}_{n-2}{\mathfrak B}^{-1}_{n-1}{\overline {\mathfrak B}}_{n-1}^{-1}
{\overline {\mathfrak B}}_{n-2}\bigr]
-N\tr\bigl[{\mathfrak B}_{n-1}{\overline {\mathfrak B}}_{n-1}\bigr]\Bigr\}.
\label{model3}
\eea

Here it becomes clear why we demand all matrices except ${\mathfrak B}_2$ to be quadratic: 
we must be able to invert them in order to write the corresponding generating function as a free energy of a chain of Hermitian matrices, as we
demonstrate below.

We now recall \cite{AKM} that we can write an integral over general complex matrices ${\mathfrak B}_i$
in terms of positive definite Hermitian matrices $X_i$ upon the variable changing
\be
X_i:={\overline {\mathfrak B}}_i {\mathfrak B}_i,\quad i=2,\dots, n-1.
\label{model4}
\ee
All the matrices $X_i$ ($i=2,\dots,n-1$) are of the same size $\gamma_3N\times \gamma_3N$. Changing the integration measure for
rectangular complex matrices introduces just a simple logarithmic term (see, e.g., \cite{AC}) and the
resulting integral becomes
\bea
&{}&\int {DX_2}_{\ge0}\cdots {DX_{n-1}}_{\ge0}
\exp\Bigl\{ N\sum_{r=1}^\infty \frac{t_r}{r}\tr\bigl[(X_2|\Lambda|^2)^r\bigr]-N\tr (X_2X_3^{-1})
-\cdots-N\tr(X_{n-2}X_{n-1}^{-1})\Bigr.
\nonumber\\
&{}&\quad\quad\Bigl.-N\tr X_{n-1}+(\gamma_2-\gamma_3)N\tr\log X_2-\gamma_2N\tr\log X_3
-\gamma_3N\tr\log (X_4\cdots X_{n-1})\Bigr\}.
\label{model5}
\eea
The logarithmic term in $X_2$ stabilizes the equilibrium distribution of eigenvalues of this matrix in the
domain of positive real numbers; in the case where $\gamma_2=\gamma_3$, we lose this term and must use the technique
of matrix models with hard walls (for a review, see, e.g., \cite{Ch06}).

Making a scaling $X_i\to X_i|\Lambda|^{-2}$ for all the integration variables, we reduce (\ref{model5}) to a more
familiar form of an integral over a chain of matrices,
\bea
&{}&\int {DX_2}_{\ge0}\cdots {DX_{n-1}}_{\ge0}
\exp\Bigl\{ N\sum_{r=1}^\infty \frac{t_r}{r}\tr(X_2^r)-N\tr (X_2X_3^{-1})
-\cdots-N\tr(X_{n-2}X_{n-1}^{-1})\Bigr.
\nonumber\\
&{}&\quad\quad\Bigl.-N\tr \bigl(X_{n-1}|\Lambda|^{-2}\bigr)+(\gamma_2-\gamma_3)N\tr\log X_2-\gamma_2N\tr\log X_3
-\gamma_3N\tr\log (X_4\cdots X_{n-1})\Bigr\}
\label{model6}
\eea
We use this expression when deriving the spectral curve equation in the next section. 
Now we proceed further expressing
integral (\ref{model6}) in terms of eigenvalues $x_i^{(k)}$ of the matrices $X_k$, $k=2,\dots,n-1$.

We apply the Mehta--Itzykson--Zuber integration formula to every term in the chain of matrices in (\ref{model6}).
Taking into account that, for instance, the integral over the unitary group for the term $e^{-N\tr X_kX_{k+1}^{-1}}$
gives
$$
\int DU e^{-N \sum_{i,j=1}^{\gamma_3N}U_{ij}x_i^{(k)}U^*_{ij}[x_j^{(k+1)}]^{-1}}
=\frac{\det_{i,j}[e^{-Nx_i^{(k)}/x^{(k+1)}_j}]}{\Delta(x^{(k)})\Delta(1/x^{(k+1)})}
$$
and that $1/\Delta(1/x^{(k+1)})=\prod_{i=1}^{\gamma_3N}[x_i^{(k+1)}]^{\gamma_3N-1} /\Delta(x^{(k+1)})$
we eventually write the expression in terms of eigenvalues of the matrices $X_k$:
\bea
&{}&\int_0^\infty \prod_{i=1}^{\gamma_3N}dx_i^{(2)}\frac{\Delta(x^{(2)})}{\Delta{\bigl(|\Lambda|^{-2}\bigr)}}
\prod_{k=3}^{n-1}\left(\prod_{i=1}^{\gamma_3N}\frac{dx_i^{(k)}}{x_i^{(k)}}\right)\times\nonumber\\
&{}&\quad\times\prod_{i=1}^{\gamma_3N}\Bigl[\bigl({x_i^{(2)}}/{x_i^{(3)}}\bigr)^{(\gamma_2-\gamma_3)N}
 e^{N\sum_{r=1}^\infty \frac{t_r}{r}(x_i^{(2)})^r-N x_i^{(2)}/x_i^{(3)}-\cdots-Nx_i^{(n-2)}/x_i^{(n-1)}
 -Nx_i^{(n-1)}|\Lambda|_i^{-2}}\Bigr]
 \label{model7}
\eea
Finally, if we introduce logarithmic quantities
$$
\varphi_i^{(r)}=\log x_i^{(r)}, \quad r=3,\dots,n-1,
$$
we can rewrite integral (\ref{model7}) in a more transparent form resembling that of the Today chain:
\bea
&{}&\int_0^\infty \prod_{i=1}^{\gamma_3N}dx_i^{(2)}\frac{\Delta(x^{(2)})}{\Delta{\bigl(|\Lambda|^{-2}\bigr)}}
\prod_{i=1}^{\gamma_3N}\biggl[\int_{-\infty}^\infty \prod_{k=3}^{n-1} d\varphi_i^{(k)}\times\biggr.
\nonumber\\
&{}&\quad\times\exp\Bigl[ N\sum_{r=1}^\infty \frac{t_r}{r}\bigl(x_i^{(2)}\bigr)^r
+(\gamma_2-\gamma_3)N\log x_i^{(2)} -(\gamma_2-\gamma_3)N \varphi_i^{(3)}
\Bigr.\nonumber\\
&{}&\quad\quad \biggl.\Bigl.-N x_i^{(2)}e^{-\varphi_i^{(3)}}-Ne^{\varphi_i^{(3)}-\varphi_i^{(4)}}-\cdots
-Ne^{\varphi_i^{(n-2)}-\varphi_i^{(n-1)}}-Ne^{\varphi_i^{(n-1)}}|\Lambda|_i^{-2}\Bigr]\biggr].
\label{model8}
\eea
In this form it is clear that
all integrals w.r.t. $\varphi_i^{(k)}$ are convergent.

\section{The case of two-profile Belyi morphism for $n=3$}\label{s:general}
\setcounter{equation}{0}

We now recall the results of \cite{AC} where the case $n=3$ was considered. In this case, we do not
have ``intermediate'' integrations over $\varphi_i$ in (\ref{model8}) and the partition function is
described by the following lemma.

\begin{lm}\label{lm:Belyi}
In the case where we allow only three ramification points: $0$, $1$, and $\infty$, the generating function
\be
{\mathcal F}[\{t_1,t_2,\dots\},\{{\mathfrak t}_1,{\mathfrak t}_2,\dots\},\beta;N]=\sum_{\Gamma}
\frac{1}{|\mathop{Aut}\Gamma|}
N^{2-2g}\beta^{n_2}\prod_{i=1}^{n_1}t_{r_i}\prod_{k=1}^{n_3}{\mathfrak t}_{s_k}
\label{gen-fun-gen}
\ee
of Belyi morphisms in which we fix two sets of ramification profiles: $\{t_{r_1},\dots,t_{r_{n_1}}\}$ at infinity
and $\{{\mathfrak t}_{s_1},\dots,{\mathfrak t}_{s_{n_3}}\}$ at $1$ and we take a sum over profiles at zero,
is given by the integral over Hermitian positive definite $(\gamma N\times \gamma N)$-matrix $X$
with the external matrix field $\tilde\Lambda:=|\Lambda|^{-2}$:
\be
{\mathcal Z}[t,\mathfrak t]=\prod_{k=1}^{\gamma N}|\lambda_k|^{-2\beta N}
\int\limits_{\gamma N\times \gamma N} DX_{\ge 0}
e^{N\tr\Bigl[-X|\Lambda|^{-2}+ \sum\limits_{m=1}^\infty \frac {t_m}m X^m+(\beta-\gamma)\log X\Bigr]}.
\label{tt2}
\ee
Here ${\mathfrak t}_{s}=\tr\bigl[(\Lambda{\overline\Lambda})^s\bigr]$.
\end{lm}

Integral (\ref{tt2}) is a GKM integral \cite{MMM}; after integration
over eigenvalues $x_k$ of the matrix $X$ it acquires the form of the ratio of two determinants,
\be
{\mathcal Z}[t,\mathfrak t]=\prod_{k=1}^{\gamma N}|\lambda_k|^{-2\beta N}
\frac{\Bigl\| \frac{\partial^{k_1-1}}{\partial {\tilde\lambda}_{k_2}^{k_1-1}}f(\tilde\lambda_{k_2})
\Bigr\|_{k_1,k_2=1}^{\gamma N}}{\Delta(\tilde\lambda)},
\label{tt3}
\ee
where
\be
f(\tilde\lambda)=\int_{0}^{\infty}x^{N(\beta-\gamma)}
e^{-Nx\tilde\lambda+N\sum\limits_{m=1}^\infty \frac {t_m}m x^m}.
\label{tt4}
\ee
Because any GKM integral (in the proper normalization) is a $\tau$-function of the
KP hierarchy, and for a model with the logarithmic term in the potential it was demonstrated in
\cite{MMS}, we immediately come to the conclusion that the exponential
$e^{{\mathcal F}[\{t\},\{{\mathfrak t}\},\gamma;N]}$
of generating function (\ref{gen-fun-gen}) modulo the
normalization factor $\prod_{k=1}^{\gamma N}|\lambda_k|^{-2\beta N}$
is a $\tau$-function of the KP hierarchy (that is, it satisfies the bilinear Hirota relations)
in times ${\mathfrak t}_s$ described in Lemma~\ref{lm:Belyi}.

\section{Spectral curve and topological recursion}\label{s:CEO}
\setcounter{equation}{0}
In this section, we propose the method for deriving the spectral curve of model (\ref{model6}) 
adapting the technique of \cite{EPF} to our case of a nonstandard interaction between matrices in the matrix chain. In
the present short paper,
we restrict ourselves to a technically more transparent case of the three-matrix model given by the integral
\be
\int DM_1\,DM_2\,DM_3\,e^{N\tr[V(M_1)+M_1M_2^{-1}-\gamma_2\log M_2+M_2M_3+U(M_3)]},
\label{three-1}
\ee
where the integrations are performed w.r.t.
positive-definite Hermitian matrices of size $\gamma_3N\times \gamma_3N$ and potentials
$V(x)$ and $U(x)$ are two Laurent polynomials of the respective positive degrees $n$ and $r$
(this consideration can be easily generalized to the case where $V'(x)$ and $U'(x)$ are
two rational functions).

The model (\ref{three-1}) satisfies \cite{HO}, \cite{HP} equations of the two-dimensional Toda chain hierarchy, not
those of the KP hierarchy, 
but these two classes of models are closely related by Miwa-type transformations, so solving the
problem of finding the spectral curve in one model can be
translated in a standard way to solving
the corresponding problem in the other model. Because finding spectral curves for multi-matrix models is 
technically somehow more transparent than finding spectral curves for models with external matrix fields, we stay with the 
first choice.

We consider the following variations of the matrix fields $M_i$:
\bea
  \delta M_1&=&\dfrac{1}{x-M_1}\xi(M_2,M_3),\nonumber \\
  \delta M_2&=&M_2\dfrac{1}{x-M_1}\eta(M_1,M_3),
  \label{three-2}
\\
  \delta M_3&=&\dfrac{1}{x-M_1}\rho(M_1,M_2), \nonumber
\eea
where we choose $\xi$, $\eta$, and $\rho$ to be
Laurent polynomials in their arguments. We introduce the standard notation
for the leading term of the $1/N^2$-expansion of the
one-loop mean of the matrix field $M_1$:
\be
\omega_1(x):=\frac 1N\left\langle\tr\frac{1}{x-M_1}\right\rangle_0.
\ee
Here and hereafter, the subscript $0$ of a correlation function indicates
the contribution of the leading order in $1/N^2$-expansion. A single trace symbol pertains
to the whole expression inside the corresponding brackets.

The exact loop equations obtained upon variations (\ref{three-2}) read
\bea
&{}&\frac{1}{N^2}\left\langle\tr\frac{1}{x-M_1}\tr \frac{1}{x-M_1}\xi(M_2,M_3)\right\rangle^{\text{c}}
+\bigl[\omega_1(x)+V'(x)\bigr]\left\langle\tr \frac{1}{x-M_1}\xi(M_2,M_3)\right\rangle\nonumber\\
&{}&\quad +\left\langle\tr\frac{V'(M_1)-V'(x)}{x-M_1}\xi(M_2,M_3)\right\rangle
+\left\langle\tr M_2^{-1}\frac{1}{x-M_1}\xi(M_2,M_3)\right\rangle=0;
\label{i}\\
&{}& \left\langle\tr \frac{-M_1}{x-M_1}\eta(M_1,M_3)M_2^{-1}\right\rangle
+\left\langle\tr M_3M_2\frac{1}{x-M_1}\eta(M_1,M_3)\right\rangle\nonumber\\
&{}&\quad +(\gamma_2-\gamma_3)\left\langle\tr \frac{1}{x-M_1}\eta(M_1,M_3)\right\rangle=0;
\label{ii}\\
&{}& \left\langle\tr M_2\frac{1}{x-M_1}\rho(M_1,M_2)\right\rangle
+\left\langle\tr U'(M_3)\frac{1}{x-M_1}\rho(M_1,M_2)\right\rangle=0.
\label{iii}
\eea
A complete information on the model is encoded in these loop equations; solving them we can develop the topological
recursion procedure for evaluating terms in the $1/N^2$-expansion. Our goal in this paper is however more modest: we
are only going to derive the spectral curve (this nevertheless ensures all the necessary ingredients of the
topological recursion \cite{Ey}, \cite{ChEy}, \cite{CEO}, see also \cite{AlMM}, 
which are the spectral curve itself and two meromorphic differentials defined on this curve).

Because we obtain the spectral curve in the large-$N$ limit, we disregard the first term in (\ref{i}), which is of
the next order in $1/N^2$. All other terms in all three equations contribute to the leading order. 

We next perform several substitutions enabling us to produce the required identities; in all the identities below we
keep only leading terms in the large-$N$ limit:

%\bea
%\rho(M_1,M_2)=M_2^{-1}:&{}&\left\langle\tr \frac{1}{x-M_1}\right\rangle
%+\left\langle\tr M_2^{-1}U'(M_3)\frac{1}{x-M_1}\right\rangle=0,\nonumber\\
%&{}& \omega_1(x)+\left\langle\tr M_2^{-1}U'(M_3)\frac{1}{x-M_1}\right\rangle_0=0;
%\label{three-3}
%\eea
%\bea
%\xi(M_2,M_3)=U'(M_3):&{}&[\omega_1(x)+V'(x)]\left\langle\tr \frac{1}{x-M_1}U'(M_3)\right\rangle_0
%\nonumber\\
%&{}&+\left\langle\tr U'(M_3)\frac{V'(M_1)-V'(x)}{x-M_1}\right\rangle+
%\left\langle\tr M_2^{-1}U'(M_3)\frac{1}{x-M_1}\right\rangle_0=0.\nonumber
%\eea
%Using (\ref{three-3}) and introducing the polynomial
%\be
%P_{n-1}:=\left\langle\tr U'(M_3)\frac{V'(M_1)-V'(x)}{x-M_1}\right\rangle
%\label{Pn-1}
%\ee
%we obtain that
%\be
%\left\langle\tr \frac{1}{x-M_1}U'(M_3))\right\rangle_0=\frac{\omega_1(x)-P_{n-1}(x)}{\omega_1(x)+V'(x)}.
%\ee

The first substitution is
\bea
&{}&\xi(M_2,M_3)=\frac{U'(M_3)-U'(z)}{M_3-z}:=\xi_0(M_3,z):\nonumber\\
&{}&\qquad \bigl[\omega_1(x)+V'(x)\bigr]\left\langle\tr \frac{1}{x-M_1}\frac{U'(M_3)-U'(z)}{M_3-z}\right\rangle_0
+\left\langle\tr \frac{U'(M_3)-U'(z)}{M_3-z}\frac{V'(M_1)-V'(x)}{x-M_1}\right\rangle_0\nonumber\\
&{}&\qquad +\left\langle\tr M_2^{-1}\frac{1}{x-M_1}\frac{U'(M_3)-U'(z)}{M_3-z}\right\rangle_0=0.
\label{three-4}
\eea
For the last term in (\ref{three-4}), we use equation (\ref{ii}):
\bea
&{}&\left\langle\tr M_2^{-1}\frac{1}{x-M_1}\frac{U'(M_3)-U'(z)}{M_3-z}\right\rangle_0
\nonumber\\
&{}&=\left\langle\tr M_2M_1^{-1}\frac{1}{x-M_1}\frac{U'(M_3)-U'(z)}{M_3-z}M_3\right\rangle_0+
(\gamma_3-\gamma_2)\left\langle\tr M_1^{-1}\frac{1}{x-M_1}\frac{U'(M_3)-U'(z)}{M_3-z}\right\rangle_0
\nonumber\\
&{}&=\frac 1x\left\langle\tr M_2\frac{1}{x-M_1}\frac{U'(M_3)-U'(z)}{M_3-z}(M_3-z+z)\right\rangle_0
+\frac 1x \left\langle\tr M_2 M_1^{-1}\frac{U'(M_3)-U'(z)}{M_3-z} M_3\right\rangle_0
\nonumber\\
&{}&\quad +(\gamma_3-\gamma_2)\frac 1x\left\langle\tr \frac{1}{x-M_1}\frac{U'(M_3)-U'(z)}{M_3-z}\right\rangle_0
+(\gamma_3-\gamma_2)\frac 1x \left\langle\tr M_1^{-1}\frac{U'(M_3)-U'(z)}{M_3-z}\right\rangle_0\nonumber\\
&{}&=\frac zx \left\langle\tr M_2\frac{1}{x-M_1}\frac{U'(M_3)-U'(z)}{M_3-z}\right\rangle_0
+\frac 1x\left\langle\tr M_2\frac{1}{x-M_1}(U'(M_3)-U'(z))\right\rangle_0\nonumber\\
&{}&\qquad\qquad 
+(\gamma_3-\gamma_2)\frac 1x\left\langle\tr \frac{1}{x-M_1}\frac{U'(M_3)-U'(z)}{M_3-z}\right\rangle_0
+\frac 1x \left\langle\tr M_2^{-1}\frac{U'(M_3)-U'(z)}{M_3-z}\right\rangle_0,
\label{three-5}
\eea
where in the last term we have again used substitution (\ref{ii}) (in opposite direction).
We introduce the polynomials
\be
\begin{array}{l}
P_{n-1,r-1}(x,z):=\left\langle\tr \dfrac{U'(M_3)-U'(z)}{M_3-z}\dfrac{V'(M_1)-V'(x)}{x-M_1}\right\rangle_0,
\phantom{\biggl|}
\\
Q_{r-1}(z):=\left\langle\tr M_2^{-1}\dfrac{U'(M_3)-U'(z)}{M_3-z}\right\rangle_0.
\phantom{\biggl|}
\end{array}
\label{three-6}
\ee
Equation (\ref{three-4}) then becomes
\bea
&{}&\Bigl[\omega_1(x)+V'(x)+\frac{\gamma_3-\gamma_2}x \Bigr]
\left\langle\tr \frac{1}{x-M_1}\frac{U'(M_3)-U'(z)}{M_3-z}\right\rangle_0
\nonumber\\
&{}&\quad
+\frac{z}{x} \left\langle\tr M_2 \frac{1}{x-M_1}\frac{U'(M_3)-U'(z)}{M_3-z}\right\rangle_0
+\frac 1x \left\langle\tr M_2\frac{1}{x-M_1}(U'(M_3)-U'(z))\right\rangle_0\nonumber\\
&{}&\quad +P_{n-1,r-1}(x,z)+\frac 1x Q_{r-1}(z)=0,
\label{three-7}
\eea
and it remains only to evaluate the term $\left\langle\tr M_2\frac{1}{x-M_1}(U'(M_3)-U'(z))\right\rangle_0$. Note
first that, from (\ref{iii}), we have that
$$
\left\langle\tr M_2\frac{1}{x-M_1}U'(M_3)\right\rangle_0=\left\langle\tr M_2^2\frac{1}{x-M_1}\right\rangle_0,
$$
and we can evaluate $\left\langle\tr M_2\frac{1}{x-M_1}\right\rangle_0$ and
$\left\langle\tr M_2^2\frac{1}{x-M_1}\right\rangle_0$ consequently substituting $\xi(M_2,M_3)=M_2$
and $\xi(M_2,M_3)=M_2^2$ in (\ref{i}). We introduce two more polynomials
\be
{\widehat P}_{n-1}(x):=\left\langle\tr \frac{V'(M_1)-V'(x)}{x-M_1}M_2\right\rangle_0\ \hbox{and}\
{\widehat{\widehat P}}_{n-1}(x):=\left\langle\tr \frac{V'(M_1)-V'(x)}{x-M_1}M_2^2\right\rangle_0.
\label{three-8}
\ee
The substitution $\xi(M_2,M_3)=M_2$ results in the equation
$$
\bigl[\omega_1(x)+V'(x)\bigr]
\left\langle\tr M_2\frac{1}{x-M_1}\right\rangle_0 +{\widehat P}_{n-1}(x)+\omega_1(x)=0,
$$
whereas the substitution $\xi(M_2,M_3)=M_2^2$ gives
$$
\bigl[\omega_1(x)+V'(x)\bigr]\left\langle\tr M_2^2\frac{1}{x-M_1}\right\rangle_0
+{\widehat{\widehat P}}_{n-1}(x)+\left\langle\tr M_2\frac{1}{x-M_1}\right\rangle_0=0,
$$
and we obtain that
\bea
\left\langle\tr M_2\frac{1}{x-M_1}\right\rangle_0&=&
-\frac{\omega_1(x)+{\widehat P}_{n-1}(x)}{\omega_1(x)+V'(x)}\label{three-11}\\
\left\langle\tr M_2^2\frac{1}{x-M_1}\right\rangle_0&=&\frac{1}{\omega_1(x)+V'(x)}
\Bigl[-{\widehat{\widehat P}}_{n-1}(x)+\frac{\omega_1(x)+{\widehat P}_{n-1}(x)}{\omega_1(x)+V'(x)}\Bigr].
\label{three-12}
\eea
Equation (\ref{three-7}) therefore takes the form
\bea
&{}&\Bigl[\omega_1(x)+V'(x)+\frac{\gamma_3-\gamma_2}x \Bigr]
\left\langle\tr \frac{1}{x-M_1}\frac{U'(M_3)-U'(z)}{M_3-z}\right\rangle_0
\nonumber\\
&{}&\quad
+\frac{z}{x} \left\langle\tr M_2 \frac{1}{x-M_1}\frac{U'(M_3)-U'(z)}{M_3-z}\right\rangle_0
+s(x,z)=0,
\label{three-13}
\eea
in which $s(x,z)$ is a rational function
\bea
s(x,z)&=&P_{n-1,r-1}(x,z)+\frac 1x Q_{r-1}(z)\nonumber\\
&{}&\quad +\frac 1x\biggl(\frac {1}{\omega_1+V'(x)}
\Bigl[-{\widehat{\widehat P}}_{n-1}(x)+\frac{\omega_1(x)+{\widehat P}_{n-1}(x)}{\omega_1(x)+V'(x)}\Bigr]
+U'(z)\frac{\omega_1(x)+{\widehat P}_{n-1}(x)}{\omega_1(x)+V'(x)}\biggr).
\label{three-14}
\eea

Performing the last substitution
$\xi(M_2,M_3)=\frac{U'(M_3)-U'(z)}{M_3-z}M_2$ in (\ref{i}), we obtain
\bea
&{}&\bigl[\omega_1(x)+V'(x)\bigr]\left\langle\tr M_2 \frac{1}{x-M_1}\frac{U'(M_3)-U'(z)}{M_3-z}\right\rangle_0
\nonumber\\
&{}&\qquad
+\left\langle\tr \frac{1}{x-M_1}\frac{U'(M_3)-U'(z)}{M_3-z}\right\rangle_0 +t(x,z)=0,
\label{three-15}
\eea
where
\be
t(x,z):={\widehat P}_{n-1,r-1}(x,z)
:=\left\langle\tr M_2\frac{U'(M_3)-U'(z)}{M_3-z}\frac{V'(M_1)-V'(x)}{x-M_1}\right\rangle_0
\label{three-16}
\ee
is again a polynomial function. We now treat Eqs. (\ref{three-13}) and (\ref{three-15}) as a system of two linear
equations on two unknowns $\left\langle\tr \frac{1}{x-M_1}\frac{U'(M_3)-U'(z)}{M_3-z}\right\rangle_0$
and $\left\langle\tr M_2\frac{1}{x-M_1}\frac{U'(M_3)-U'(z)}{M_3-z}\right\rangle_0$. We are interested
in the case where this system is degenerate, which imposes the constraint on the variable $z$:
\be
\det\left[
      \begin{array}{cc}
        \omega_1(x)+V'(x)+\frac{\gamma_3-\gamma_2}x & z/x \\
        1 & \omega_1(x)+V'(x) \\
      \end{array}
    \right]=0,
\ee
which gives
\be
z=x(\omega_1(x)+V'(x))\Bigl(\omega_1(x)+V'(x)+\frac{\gamma_3-\gamma_2}x\Bigr).
\label{zy}
\ee
It is a standard trick in multi-matrix models to introduce the new variable $y$:
\be
y:=\omega_1(x)+V'(x).
\label{y}
\ee
Then the condition of solvability of the system of  linear equations (\ref{three-13}) and (\ref{three-15}) is exactly the
\emph{spectral curve equation}
\be
s(x,z)-\Bigl(y+\frac{\gamma_3-\gamma_2}x\Bigr)t(x,z)=0,\quad\hbox{where}\quad z=xy^2+(\gamma_3-\gamma_2)y.
\label{spectral}
\ee

Despite its complexity even in the simplest cases (say, we obtain a hyperelliptic
curve of maximum genus three for the Gaussian potentials $V(x)$ and $U(z)$ in the Example~\ref{ex:1} below),
we still have an algebraic curve in contrast to the case of Hurwitz numbers in the case of branching points with
only simple ramifications for which it was conjectured in \cite{Marino} and shown in \cite{EyMar} that the corresponding spectral curve 
in the case of simple Hurwitz numbers
is the Lambert curve given by a nonpolynomial equation $x=ye^{-y}$.

\begin{example}\label{ex:1}
Let us consider the case of Gaussian potentials $V(x)=x^2/2$ and $U(z)=z^2/2$. Then all the polynomials
${P}_{n-1,r-1}$, ${\widehat P}_{n-1,r-1}$, ${\widehat P}_{n-1}$, ${\widehat {\widehat P}}_{n-1}$, and $Q_{r-1}$
are constants and, moreover, ${P}_{n-1,r-1}=1$ and ${\widehat P}_{n-1,r-1}={\widehat P}_{n-1}$.
Then, after all cancelations, we obtain the spectral curve equation
\be
y-x+{\widehat P}-{\widehat{\widehat P}}y+xy^2+Qy^2+y^2(y-x)(xy+\gamma_2-\gamma_3)=0,
\label{sp}
\ee
which, for the general values of constants in (\ref{sp}), describes a hyperelliptic curve of genus three.
\end{example}

\section{Conclusion}
We have constructed the chain of matrix representation for the
generating functions for numbers of generalized Belyi fat graphs for hypergeometric
Hurwitz numbers with ramifications at $n$ distinct points and with ramification profiles fixed at two of these
$n$ points. We also distinguish between fat graphs with different numbers of pre-images of other ramification points.
The corresponding partition functions lie
in the generalized Kontsevich matrix-model class thus being tau functions of the KP hierarchy, which was 
previously shown from the character expansion standpoint in \cite{HO}.
We were able to construct the chain of matrix representation with a nonstandard interaction
$\sum_{i=3}^n \tr(M_{i-1}M_i^{-1})$ between neighbor Hermitian positive-definite matrices in the chain
in the case where we do not distinguish between variables of $n-3$ cycles. We were successful in proposing
a method for solving models with interactions of this sort. For the simplicity sake, in this note we have restricted our consideration
to the model case of the two-dimensional Toda chain hierarchy with one intermediate matrix (the case $n=4$), but our method 
can be straightforwardly generalized to the case of $n-3$ intermediate matrices with the 
last, $n$th matrix, being an external field $|\Lambda|^{-2}$. 

It is interesting to establish other relations. For instance, generating function (\ref{gen-fun})
in the case of clean Belyi morphisms is related \cite{AC} to the 
free energy of the Kontsevich--Penner matrix model \cite{ChM}, \cite{ChM2}, which is known (see
\cite{Ch95},\cite{Norbury},\cite{DoNor}) to be the generating function of
the numbers of integer points in moduli spaces
${\mathcal M}_{g,s}$ of curves of genus $g$ with $s$ holes with fixed (integer) perimeters; the very same model
was also related \cite{Ch95} by a canonical transformation to two copies of the Kontsevich matrix model expressed
in times related to the discretization of the moduli spaces ${\mathcal M}_{g,s}$. It is tempting to generalize
these discretization patterns to cut-and-join operators of \cite{Zograf} and \cite{AMMN} in the case of
hypergeometric Hurwitz numbers and to Hodge integrals of \cite{Kazarian}.

\section*{Acknowledgments}
it is our pleasure to celebrate with this paper the 75th birthday of Andrey Alekseevich Slavnov, an outstanding scientist, a great personality, and the teacher of one of us (L.Ch.).

The authors acknowledge support from the ERC Advance Grant 291092 ``Exploring the Quantum Universe'' (EQU).
J.A. acknowledges support of the FNU, the Free Danish Research Council, from the grant ``Quantum gravity and the
role of black holes.'' The work of L.Ch. was supported by the Russian Foundation for Basic Research
(Grant Nos. 14-01-00860-a and 13-01-12405-ofi-m) and by the Program 19-P of the Russian Academy of Sciences ``Fundamental Problems of Nonlinear Dynamics.''

\def\thetheorem{\Alph{section}.\arabic{theorem}}
\def\theprop{\Alph{section}.\arabic{prop}}
\def\thelemma{\Alph{section}.\arabic{lm}}
\def\thecor{\Alph{section}.\arabic{cor}}
\def\theexam{\Alph{section}.\arabic{exam}}
\def\theremark{\Alph{section}.\arabic{remark}}
\def\theequation{\Alph{section}.\arabic{equation}}

\setcounter{section}{0}

%\appendix{Deriving the Jacobian of transformation (\ref{RR2UVM})}\label{se:notation}
%\setcounter{equation}{0}


\begin{thebibliography}{99}

\footnotesize\itemsep=0pt

\bibitem{AlMM}
A.~Alexandrov, A.~Mironov and A.~Morozov, Partition functions of matrix models as the first special functions of String Theory I. Finite size Hermitean 1-matrix model, {\sl Int. J. Mod. Phys.} {\bf A19} (2004) 4127-4165.

\bibitem{AMMN}
A.~Alexandrov, A.~Mironov, A.~Morozov, and S.~Natanzon, On KP-integrable Hurwitz functions, arXiv:1405.1395

\bibitem{AC}
J. Ambj{\o}rn and L. Chekhov, The matrix model for dessins d'enfants, {Ann. Inst. Henri Poincar\'e, Comb. Phys. Interact.} {\bf1} (2014) 337--361; DOI 10.4171/AIHPD/10.

%\bibitem{ACKM}
%J. Ambj{\o}rn, L. Chekhov, C.F. Kristjansen, and Yu. Makeenko,
%Matrix model calculations beyond the spherical limit,
%{\sl Nucl.Phys.} {\bf B404} (1993) 127-172; Erratum-ibid. {\bf B449} (1995) 681; hep-th/9302014.

%\bibitem{ACM} J. Ambj{\o}rn, L. Chekhov, and Yu. Makeenko,
%Higher genus correlators and W infinity from the hermitian
%one matrix model, {\sl Phys. Lett.} {\bf B282} (1992) 341-348; hep-th/9203009


\bibitem{AKM}
J. Ambj{\o}rn, C.~F.~Kristjansen, and Y.~M.~Makeenko, Higher genus correlators for the complex matrix model,
{\sl Mod. Phys. Lett.} {\bf A7} (1992) 3187-3203; hep-th/9207020.

\bibitem{Belyi} G.~Belyi, On Galois extension of a maximal cyclotomic field,
{\sl USSR Math. Izvestiya} {\bf14}:2 (1980) 247--256

\bibitem{EyMar}
G. Borot, B. Eynard, M. Mulase, and B. Safnuk, A matrix model for simple Hurwitz numbers, and topological recursion,
{\sl J. Geom. Phys.} {\bf 61(2)} (2011) 522--540; arXiv:0906.1206. 

\bibitem{Marino}
V. Bouchard and M. Mari{\~n}o, Hurwitz numbers, matrix models, and enumerative geometry,
in:  From Hodge Theory to Integrability and tQFT: tt*-geometry, Proceedings of Symposia in Pure Mathematics, AMS (2008); arXiv:0709.1458.

%\bibitem{BIPZ} E.~Br\'ezin, C.~Itzykson, G.~Parisi, and J.-B. Zuber, Planar diagrams,
%{\sl Commun. Math. Phys.} {\bf59} (1978) 35-55.

%\bibitem{Ch} L. Chekhov, Matrix models solutions in $1/N$-expansion,
%{\sl Russ. Math. Surv.}, 61(3), 2006, pp. 1--61.

\bibitem{Ch06} L.Chekhov, Matrix models with hard walls: Geometry and
solutions, {\sl J. Phys. A} {\bf 39} (2006) 8857-8894; hep-th/0602013.

\bibitem{Ch95} L. Chekhov, Matrix models tools and geometry of moduli spaces,
{\sl Acta Appl. Mathematicae} {\bf 48} (1997) 33-90; e-Print Archive:
hep-th/9509001.

\bibitem{ChEy} L. Chekhov, B. Eynard,
Hermitean matrix model free energy: Feynman graph technique
for all genera, {\sl JHEP} {\bf 0603}:014
(2006); hep-th/0504116.

\bibitem{CEO} L. Chekhov, B. Eynard, and N. Orantin,
Free energy topological expansion for the 2-matrix model,
{\sl JHEP} {\bf 12}(2006)053; hep-th/0603003.

\bibitem{ChM} L. Chekhov and Yu. Makeenko, The multicritical Kontsevich-Penner model,
{\sl Mod. Phys. Lett.} {\bf A7} (1992) 1223-1236; hep-th/9201033.

\bibitem{ChM2} L. Chekhov and Yu. Makeenko,
A hint on the external field problem for matrix models,
{\sl Phys. Lett.} {\bf B278} (1992) 271-278; hep-th/9202006

%\bibitem{ChPal} L. Chekhov and K.~Palamarchuk, Two logarithm matrix model with an external field,
%{\sl Mod. Phys. Lett.} {\bf A14} (1999) 2229-2244; e-Print Archive: hep-th/9811200

\bibitem{DoNor} Norman Do and Paul Norbury, Pruned Hurwitz numbers, ArXiv:1312.7561v1.

\bibitem{DMSA} O.~Dumitrescu, M.~Mulase, B.~Safnuk, and A.~Sorkin, The spectral curve of the Eynard--Orantin
recursion via the Laplace transform, ArXiv:1202.1159.

\bibitem{Ey} B. Eynard, All genus correlation functions for the hermitian 1-matrix model,
{\sl JHEP} {\bf 0411:031} (2004).

\bibitem{EPF} B. Eynard and A. Prats Ferrer, Topological expansion of the chain of matrices,
{\sl JHEP} {\bf 0907} (2009) 096; ArXiv:0805.1368v2.

\bibitem{GJ}
I.~P.~Goulden and D.~M.~Jackson, The KP hierarchy, branched covers, and triangulations, arXiv:0803.3980.

\bibitem{Grot} A.~Grothendieck, Esquisse d'un programme, {\sl Geometric Galois Action},
Cambridge Univ. Press, Cambridge (1997) 5--48.

\bibitem{HP} M. Guay-Paquet and J. Harnad, Generating functions for weighted Hurwitz numbers, arXiv:1408.6766.


\bibitem{HO} J. Harnad and A. Yu. Orlov, Hypergeometric $\tau$-functions, Hurwitz numbers
and enumeration of paths, ArXiv:1407.7800.

\bibitem{Kazarian}
M. Kazarian, KP hierarchy for Hodge integrals, {\sl Adv. Math.} {\bf 221} (2009) 1-21.

\bibitem{KZ}
M. Kazarian and P. Zograf, Virasoro constraints and topological recursion for Grothendieck?s dessin counting, arXiv:1406.5976.

\bibitem{MMM}
S. Kharchev, A.Marshakov, A.Mironov, A.Morozov, and A. Zabrodin, Unification of all string models
with $c<1$, {\sl Phys.Lett.}, {\bf 275B} (1992) 311-314.\\
S. Kharchev, A.Marshakov, A.Mironov, A.Morozov, and A. Zabrodin, Towards unified theory of 2d gravity,
{\sl Nucl. Phys.}, {\bf B380} (1992) 181-240.

%\bibitem{Migdal} A. A. Migdal, Loop equations and $1/N$ expansion, {\sl Phys. Rep.} {\bf102} (1983) 199-290.

\bibitem{MMS} A.~Mironov, A.~Morozov, and G.~Semenoff, Unitary matrix integrals in the framework of
Generalized Kontsevich Model. I. Brez\'in--Gross--Witten model,
{\sl Int. J. Mod. Phys.} {\bf A10} (1995) 2015--2040.

\bibitem{Norbury} P. Norbury, Counting lattice points in the moduli space of curves, {\sl Math. Res. Lett.}
{\bf 17} (2010) 467-481.

\bibitem{OP} A.~Okounkov and R.~Pandharipande, Gromov--Witten theory, Hurwitz numbers, and completed cycles,
{\sl Ann. Math.} {\bf163} (2006) 517-590; math.AG/0204305.

\bibitem{OS} A. Orlov and D.~M.~Shcherbin, Hypergeometric solutions of soliton equations, {\sl Theor. Math. Phys.}
{\bf128} (2001) 906--926

\bibitem{Or} A. Orlov, Hypergeometric functions as infinite-soliton tau functions, {\sl Theor. Math. Phys.}
{\bf146} (2006) 183--206.

\bibitem{Zograf} P.~G.~Zograf, Enumeration of Grothendieck's dessins and KP hierarchy, arXiv:1312:2538v2.

\end{thebibliography}
\end{document}